\documentstyle[11pt,paspconf,epsf]{article}

\begin{document}

\title{SHARC 350 \micron\ Mapping of the Galactic Center from the Caltech
Submillimeter Observatory}

\author{C. D. Dowell, D. C. Lis, E. Serabyn, M. Gardner, A. Kovacs,
\& S. Yamashita}
\affil{Dept. Physics, Caltech, Mail Code 320-47, Pasadena, CA  91125\\
cdd@submm.caltech.edu}

\begin{abstract}

Using the 20-pixel camera SHARC, we have surveyed the 350 \micron\ emission from
a 60\arcmin\ by 12\arcmin\ region in the Galactic center.  A comparable region
has been observed at 800 \micron\ by Lis \& Carlstrom
(1994); the SHARC map has better spatial resolution and sensitivity to
extended emission, however.
This paper introduces several features not visible in prior maps of dust
emission.
We have detected a 10+ pc band of emission peaking at $l = -0.14\deg,
b = +0.02\deg$, which is probably associated with the 'negative velocity arc'
observed in $^{13}$CO (Bally et al. 1988).
We suggest an association of a rounded 350 \micron\ feature near the Radio
Arc with an
expanding shell observed in CS (Tsuboi, Ukita, \& Handa 1997).
Faint emission at the same Galactic longitude as the Dust Ridge (Lis \&
Carlstrom 1994) but at opposite latitude is visible as well as a compact source
at the base of a thermal radio filament at negative Galactic latitude.
Diffuse dust emission within a few arcminutes of
the Sgr D core is detected for the first time.
350 \micron\ sources are associated with
both H$_2$O masers in Sgr D, but only one out of four OH masers
(Mehringer et al.  1998).
The mean 350 \micron /800 \micron\ flux ratio
for the Galactic center is approximately 17
($\beta \sim 2.0$) over the map but is higher ($\beta \sim 2.5$)
in parts of the Dust Ridge and lower in Sgr B2 (N).

\end{abstract}

\index{Source!Sgr A}
\index{Source!Sgr B1}
\index{Source!Sgr B2}
\index{Source!Sgr D}
\index{Dust emission}
\index{Molecular clouds}
\index{Submillimeter continuum}

\section{Introduction}

Submillimeter continuum observations of the Galactic center provide a measure
of the amount and structure of dense material on scales of 0.5 - 200 pc.
Broadband observations at 350 \micron\ suffer very little from contamination by
molecular line, free-free, and synchrotron emission, and, except toward the Sgr
B2 core, the emission is optically thin.

Lis \& Carlstrom (1994) mapped a large portion of the Galactic center at 800
\micron\ using a single bolometer and an 'on-the-fly' mapping technique; many
sources were detected for the first time, including cold clouds not seen
in the far-IR ($\lambda$ $<$ 100 \micron).  However, the beam size of the
experiment (30\arcsec) was larger than can be obtained at shorter
wavelengths, and the observing time per unit solid angle
was small.  These shortcomings have been addressed by mapping
the region at 350 \micron\ with SHARC, a multi-pixel bolometer camera.

\begin{figure}
\plotone{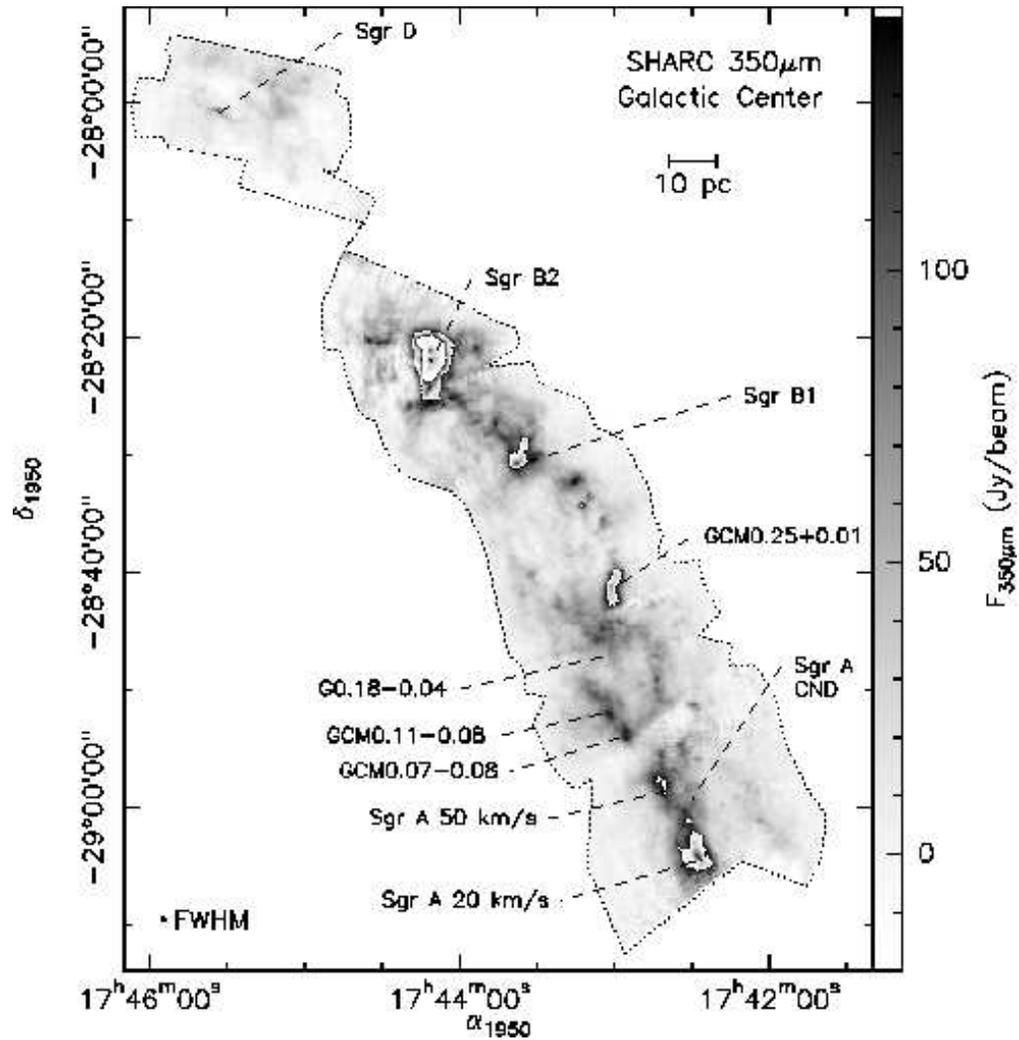}
\caption{350 \micron\ map of the Galactic center.  The beam
size is 15\arcsec .  The intensity scale wraps around and becomes nonlinear
above 145 Jy/beam.  For position reference, Sgr A$^*$ is located inside the
Circum-Nuclear Disk ('CND'), and the nonthermal radio filaments (not shown)
cross the Galactic plane near the Sickle (G0.18--0.04).}
\label{fig-a}
\end{figure}

\section{SHARC Observations}

SHARC (Wang et al. 1996) uses a linear array of 20 bolometers.  Each pixel
subtends 5\arcsec\ by 10\arcsec .  The diffraction beam
and choice of smoothing algorithm lead to an effective resolution of
15\arcsec\ FWHM in our map.

The data were acquired by 'on-the-fly' mapping (chopping and continuous
scanning) and restored with the NOD2 algorithm (Haslam 1974).
The observations took place during April-September 1998.
During the very dry conditions in April, the point-source
sensitivity per pixel was $\sim$2 Jy s$^{0.5}$ at Galactic center transit.
However,
due to map reconstruction noise and correlated sky noise (causing striping),
the sensitivity in the raw resultant map is degraded.  Fourier filtering was
applied to the map to remove the sky noise stripes on rapid timescales
(spatial wavelengths $\le$ 30\arcsec); the
filtering was quite satisfactory in that source structure was preserved.
After filtering, the typical small-scale noise was 6 Jy rms over most of
the map.

The SHARC map is shown in Figure~\ref{fig-a}.  The coverage is
approximately 0.3 square degrees.
Some imperfections are evident in the map.  There are reconstruction
artifacts to the east and west of the bright peaks of Sgr B2, a few
regions of negative flux caused by emission at the ends of the scans, and
signal-to-noise variations.

There is very good correspondence between the features in our 350 \micron\ map,
the 800 \micron\ map of Lis \& Carlstrom (1994), and the 1.3 mm map of Mezger,
Duschl, \& Zylka (1996) where coverage and sensitivity permit comparison.  In
general, we confirm the poor association of 350 \micron\ dust emission with
mid- and far-infrared ($\lambda$ $<$ 100 \micron) and radio continuum
emission.  The 350 \micron\ map resembles maps of some molecular
transitions, however (NH$_3$ -- G\"usten, Walmsley, \& Pauls 1981; CS -- Serabyn
\& G\"usten 1987).

Calibration was accomplished by observing Uranus, assuming
a temperature of 64 K (Griffin \& Orton 1993).
The map calibration assumes point-like
sources, and map fluxes are given as Jy/15\arcsec\ beam.  The calibration
uncertainty is 25\%.

\section{Discussion}

\subsection{Source Fluxes}

The source fluxes measured in our 350 \micron\ map are given in
Table~\ref{tbl-a}.
Assuming a dust temperature of 30 K (Lis \& Carlstrom 1994), a grain
emissivity of Q(350 \micron) = 1.9 x 10$^{-4}$, a grain density of 3 g/cm$^3$,
a grain radius of 0.1 \micron , a dust-to-gas density ratio of 0.01, and a
distance of 8.5 kpc, the conversion from flux to gas
mass is approximately 20 M$_{\sun}$/Jy (Hildebrand 1983).  The total mass
detected in our map is then 4.3 x 10$^{6}$ M$_{\sun}$.  By using an average
temperature of 30 K we
underestimate the mass in cold regions (GCM0.25+0.01, T = 18 K;
Lis \& Menten 1988) and overestimate the mass in hot regions (Sgr A CND,
T $\ge$ 40K; Davidson et al. 1992).

\begin{table}
\caption{Measured 350 \micron\ Fluxes (Jy)} \label{tbl-a}
\begin{center}%\scriptsize
\begin{tabular}{lrrr}
Source & 15\arcsec\ beam & 30\arcsec\ beam & total \\
\tableline
20 km/s cloud (GCM--0.13--0.08) & 294 & 640 & 16,000 \\
50 km/s cloud (GCM--0.02--0.07) & 166 & 370 & 8000 \\
CSO--0.14+0.02 & 73 & 148 & 3100 \\
Sgr A C.N.D. & 97 & 220 & 1500 \\
GCM0.07--0.08/GCM0.11--0.08 & 142 & 319 & 13,000 \\
GCM0.25+0.01 & 236 & 523 & 5900 \\
Sgr B1 & 290 & 625 & 9300 \\
Sgr B2 (M) & 2740 & 4740 & 49,000 \\
Sgr B2 (N) & 2350 & 3810 & \\
Sgr D & 106 & 197 & 7600 \\
\tableline
Total & & & 210,000 \\
\end{tabular}
\end{center}
\end{table}

\subsection{Sgr A}

\begin{figure}
\plotone{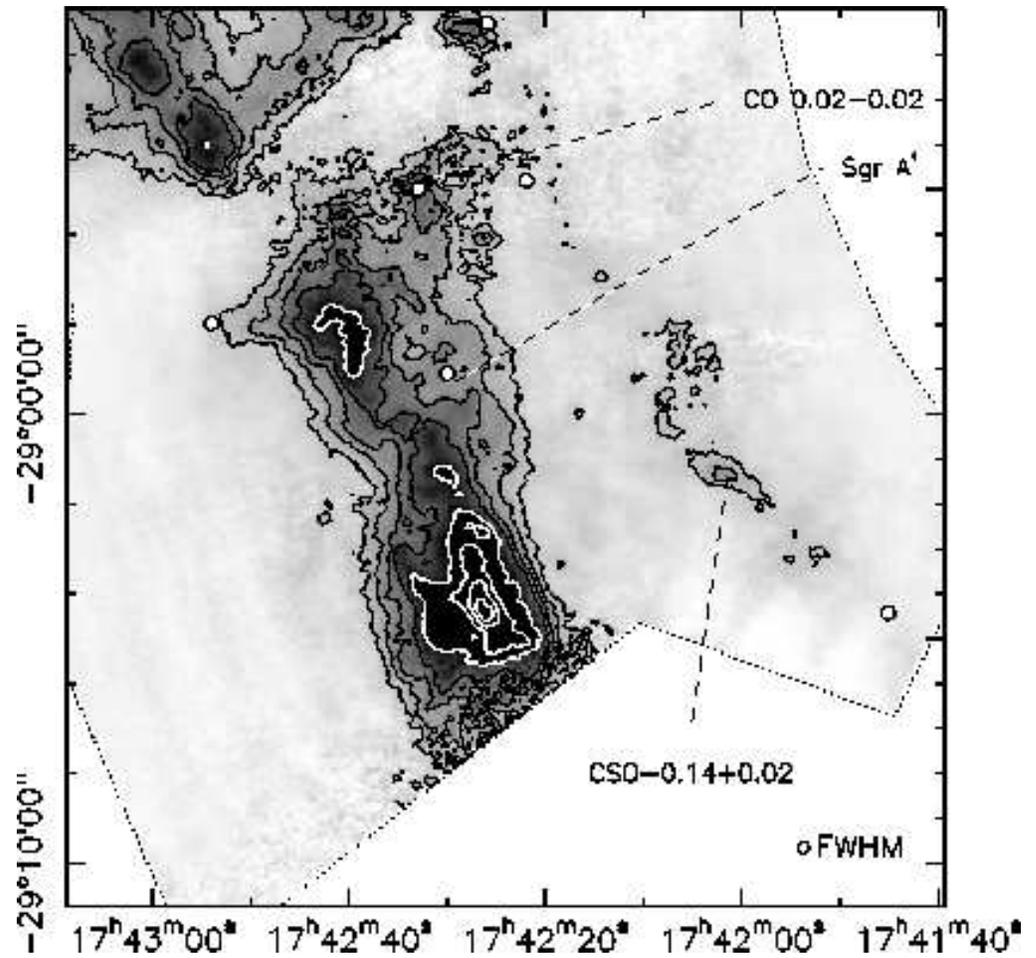}
\caption{350 \micron\ map of Sagittarius A.  The contour levels are
40, 60, 80, 100, 140, 180, 220, and 260 Jy/beam.  The brightest submillimeter
emission is from the 20 km/s and 50 km/s clouds.  The fainter Circum-Nuclear
Disk is visible partially surrounding Sgr A$^*$.  Far-IR sources (Odenwald
\& Fazio 1984) are labeled with circles.}
\label{fig-e}
\end{figure}

The (sub-)millimeter appearance of the molecular clouds in the central
10\arcmin\ of the Galaxy is well known.
The SHARC image (Figure~\ref{fig-e}) of the bright sources GCM--0.13--0.08 (+20
km/s cloud) and GCM--0.02--0.07 (+50 km/s cloud)
reproduces many features observed before at longer
wavelengths (Mezger et al. 1996; Lis \& Carlstrom 1994; Dent et al. 1993).

The emission from the vicinity of Sgr A* is relatively weak at 350 \micron .
Sgr A* itself, having a relatively flat submillimeter spectrum (Serabyn
et al. 1997), is swamped by dust emission from the 2 pc Circum-Nuclear Disk
(CND).  The CND, the dominant emitter in the central 10 pc at
$\lambda \sim 100$\micron\ (Odenwald \& Fazio 1984; Davidson et al. 1992), is
not nearly as prominent in the submillimeter, indicating a lack of cold dust
along the line of sight.

An elongated emission feature is visible to the west of GCM--0.13--0.08 running
approximately parallel to the Galactic plane.  We have labeled the peak of this
band as CSO--0.14+0.02.  The location and extent of the feature suggest an
association with the 'negative velocity arc' observed in $^{13}$CO at v = --100
to --40 km/s (Bally et
al. 1988), although the dust emission appears to be at a slightly more
negative Galactic latitude.  A far-IR source (Odenwald \& Fazio 1984) lies
along the arc, so star formation may be occuring in the cloud.

Oka et al. (1999) have recently identified a compact cloud (CO 0.02--0.02)
about 5\arcmin\ from Sgr A$^*$ with a large velocity width ($\ge$100 km/s).
Dust emission in the vicinity of this source is seen in the SHARC map
(Figure~\ref{fig-e}) as well as the maps by Dent et al. (1993), Lis \&
Carlstrom (1994), and Mezger et al. (1996).

\subsection{Radio Arc}

\begin{figure}
\plotone{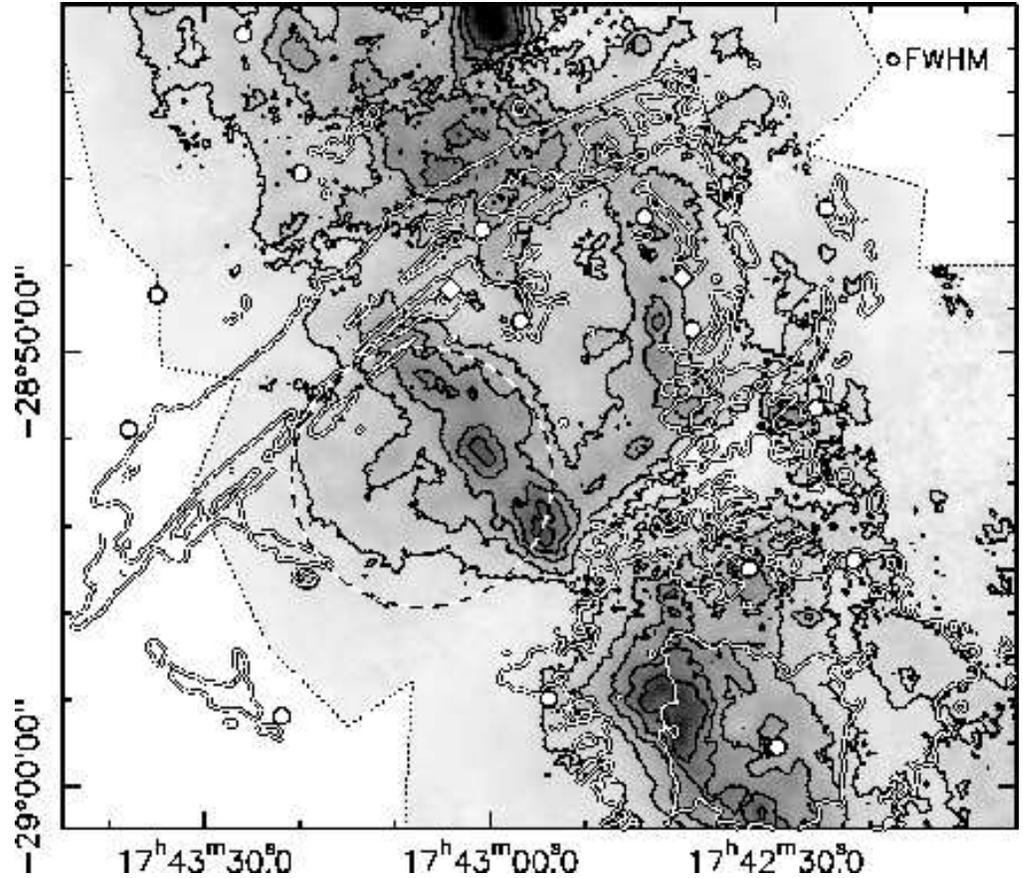}
\caption{SHARC map of the Radio Arc region.  The grayscale and single
contours show the 350 \micron\ emission with levels of 25, 50, 75, 100, and
125 Jy/beam.  The double contours show 20 cm continuum emission (Yusef-Zadeh
1986).  The system of Thermal Arched Filaments -- located at $17^h42.5^m$,
$-28\deg 49\arcmin$ -- lies to the west of a ridge of dust emission while the
radio emission from the Sickle (G0.18--0.04) at $17^h43.0^m$,
$-28\deg 47\arcmin$ is southeast of a dense cloud.
The diamonds mark known clusters of hot stars that may
ionize the Filaments and Sickle (Cotera et al. 1996).
The partial radio contour at lower left shows two thermal radio filaments
at negative Galactic latitude.  One of the filaments lies near a compact radio
and 350 \micron\ continuum source at $17^h43^m19.2^s$, $-28\deg 55\arcmin
16\arcsec$.  Its association with the filament is unknown.
The dashed circle shows the rough location of an expanding CS shell discovered
by Tsuboi et al. (1997).  The shell may also be detected at 350 \micron , as
suggested by the rounded outer contour, superposed on the ridge formed by
GCM0.07--0.08 and GCM0.11--0.08.  Circles mark far-IR sources
(Odenwald \& Fazio 1984).}
\label{fig-b}
\end{figure}

Among the prominent 20 cm features in the Galactic center are the Thermal
Arched Filaments at $l = 0.10, b = 0.05$ (Yusef-Zadeh 1986).  The SHARC map
of this region is shown in Figure~\ref{fig-b}.
The eastern filament has much more 350 \micron\ emission associated with
it than the western filament.  The majority of the 350 \micron\ flux is
east of the eastern filament, displaced further from the radio features
than is the far-IR (50 - 90 \micron ) emission (Morris, Davidson, \& Werner
1995), implying an east (cold) to west (hot) temperature gradient.  
The 'Arches' star cluster (Cotera et al. 1996; Serabyn, Shupe, \& Figer 1998)
does not seem well placed to produce this temperature gradient; a heating
source further to the west seems to be required.

A fainter thermal radio filament is located on the opposite side of the
Galactic plane, possibly associated with a compact radio feature at
$l = 0.100, b = -0.168$.  A compact 350 \micron\ source is detected in this
region with a flux of 49 Jy/15\arcsec\ beam, suggesting
recent or ongoing star formation.

Oka et al. (1998) find that at 17\arcsec\ resolution, the CO emission from
the Galactic center is composed of numerous filaments, arcs and shells.  One
of the shells, discovered in CS by Tsuboi et al. (1997), may also be seen
in the SHARC map.

\subsection{Dust Ridge}

\begin{figure}
\plotone{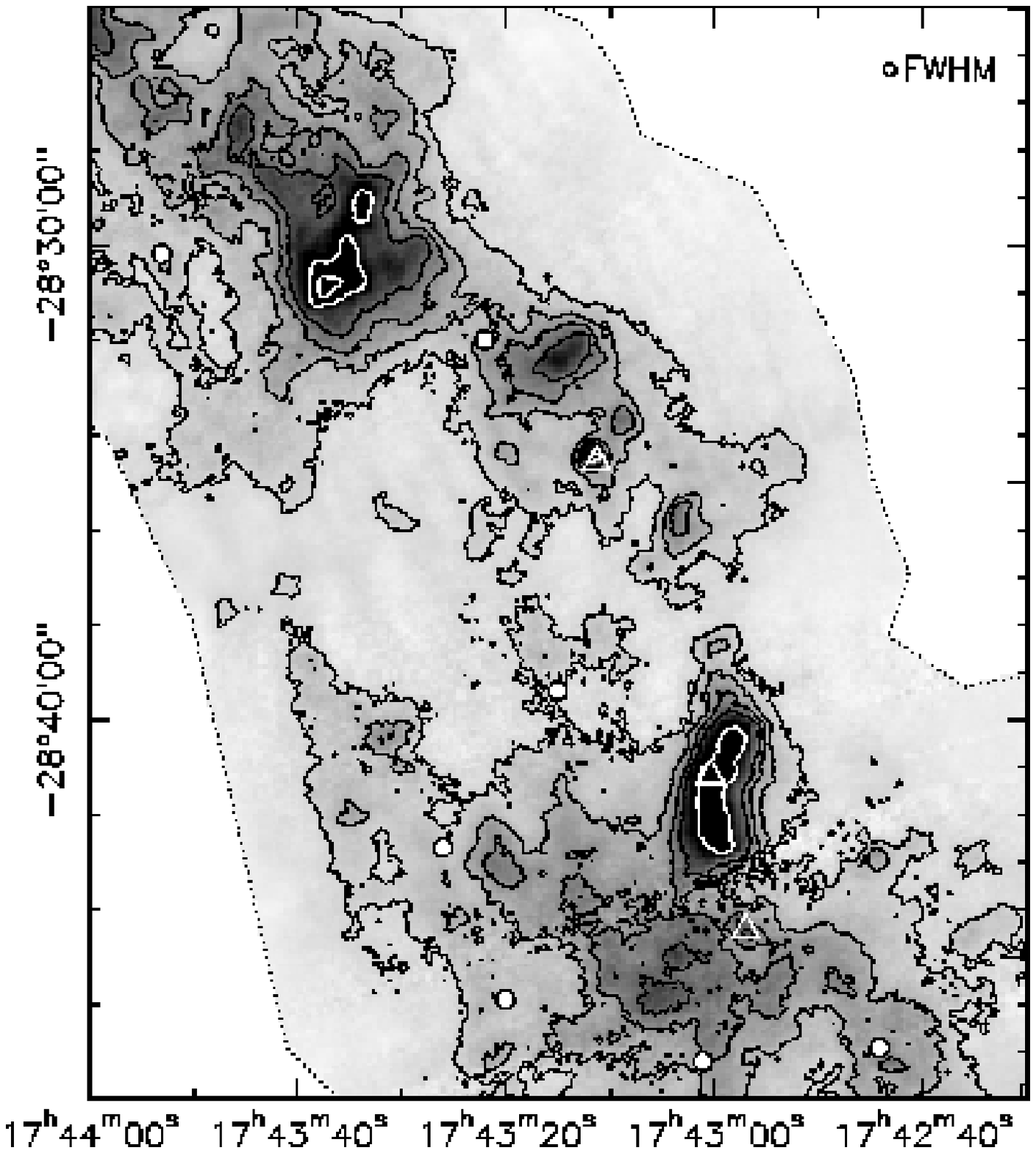}
\caption{350 \micron\ map of Dust Ridge.  Contour levels are 25, 50, 75, 100,
175, and 250 Jy/beam.  The Dust Ridge lies outside the Radio Arc
and consists of a chain of
sources extending from the bright elongated cloud GCM0.25+0.01 to Sgr B1.  A
fainter ridge of sources is located at negative Galactic latitude (eastern
part of the map).  Three H$_2$O masers (triangles) have been detected by
Lis \& Menten (1995); however, the star formation rate in the Dust Ridge as a
whole is low.  FIR sources (Odenwald \& Fazio 1984) are labeled with circles.}
\label{fig-g}
\end{figure}

Lis \& Carlstrom (1994) discovered a cold ridge of submillimeter emission
extending from GCM0.25+0.10 near the Radio Arc to Sgr B1.
The ridge is seen in absorption
against the general Galactic center emission at $\lambda$ $\la$ 70 \micron\
(Egan et al. 1998; Lis \& Menten 1998).  The SHARC 350 \micron\ map of
the Dust Ridge is shown in Figure~\ref{fig-g}.  A fainter ridge of emission
is located at negative Galactic latitude, including an extended source
at $17^h43^m31.7^s$, $-28\deg 40\arcmin 32\arcsec$ ($l = 0.330, b =
-0.076$) with a peak flux of 77 Jy/15\arcsec\ beam.

\subsection{Sgr D}

\begin{figure}
\plotone{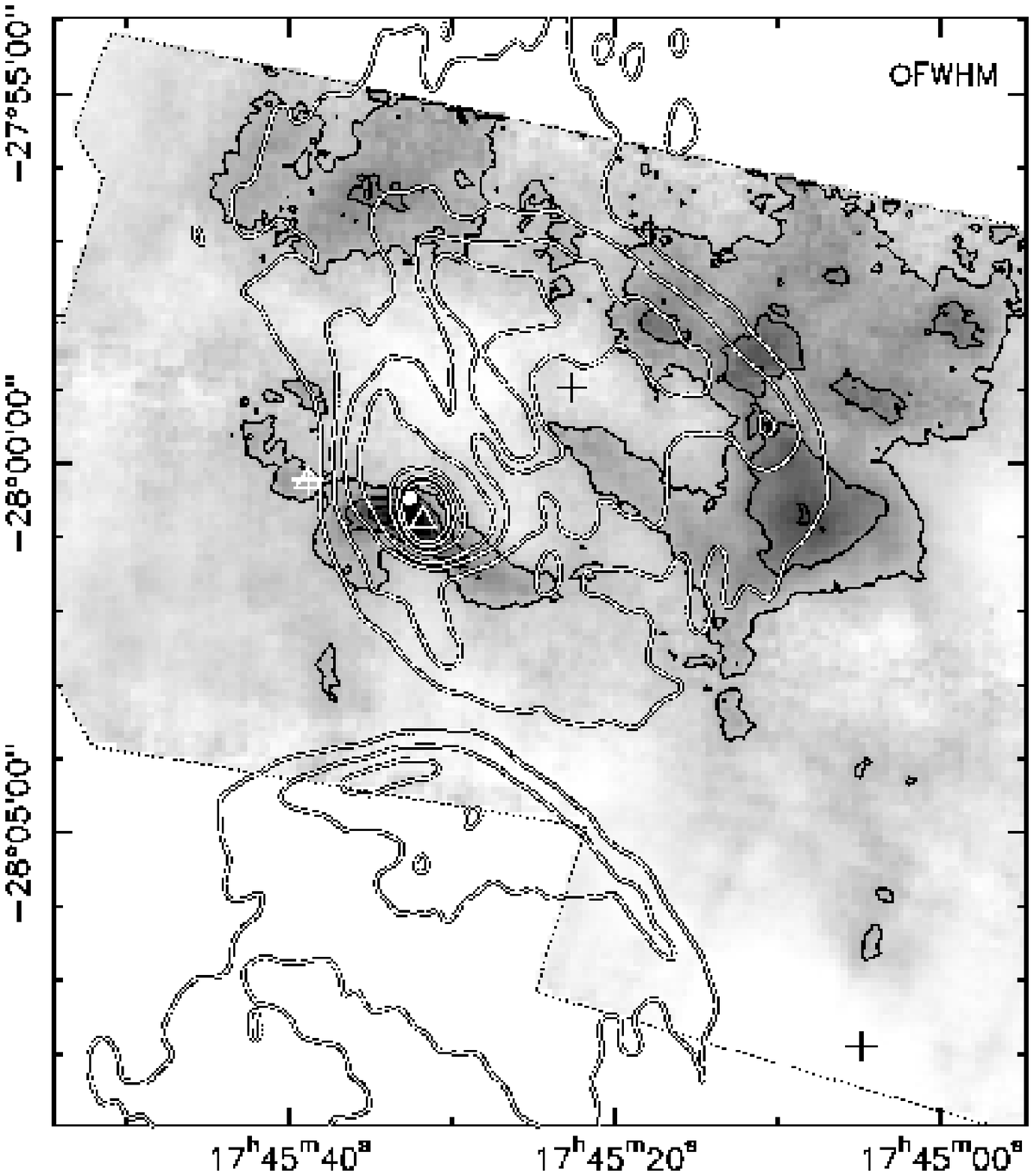}
\caption{SHARC map of Sagittarius D.  The grayscale and single contours
show the 350 \micron\ emission with levels of 20, 40, and 60 Jy/beam.
The double contours show 18 cm
continuum measured by Mehringer et al. (1998).
The triangles show the positions of H$_2$O masers, and the
plus symbols show OH masers (Mehringer et al. 1998).  The H$_2$O masers
correspond more closely with regions of submillimeter emission.  The diffuse
emission to the northwest of the Sgr D core lies along the boundary of the
H{\sc ii} region.  The FIR peak from Odenwald \& Fazio (1984) is marked with a
circle.}
\label{fig-d}
\end{figure}

Sgr D is a site of active star formation located at $l=1.13, b=-0.10$.  The
projected distance from the Galactic center is 170 pc; however, Sgr D
may lie beyond the Galactic center region (Mehringer et al. 1998).  The
800 \micron\ emission from the Sgr D core has previously been mapped by Lis,
Carlstrom, \& Keene (1991).  With SHARC, we have mapped a much larger area and
detected diffuse emission over much of the region.  The 350 \micron\ map is
compared with the 18 cm map of Mehringer et al. (1998) in Figure~\ref{fig-d}.
The peak flux at the Sgr D core is 106 Jy.  To the
northwest of the core, the submillimeter emission appears to lie along the
boundary of the H{\sc ii} region.  Perhaps the radio H{\sc ii} region is
density bounded at this position and the submillimeter emission comes from the
H{\sc ii}-molecular cloud interface.

Submillimeter emission is associated with both 1.3 cm H$_2$O masers known in the
region (Mehringer et al. 1998).  Maser source 1 is coincident with the bright
Sgr D core, while maser source 2 is associated with a modest (34 Jy/beam)
submillimeter source.
On the other hand, correspondence of 350 \micron\ sources with 18 cm OH maser
sources is lacking.  Of the 4 OH maser sources falling within our map
(E, G, H, and
I -- Mehringer et al. 1998), only one of them (I, coincident with H$_2$O
maser 2) has associated submillimeter emission.  This supports the
suspicion by Mehringer et al. (1998) that sources E, G, and H are
evolved stars rather than young stellar objects.

\subsection{350 \micron\ to 800 \micron\ Flux Ratio}

\begin{figure}
\plotone{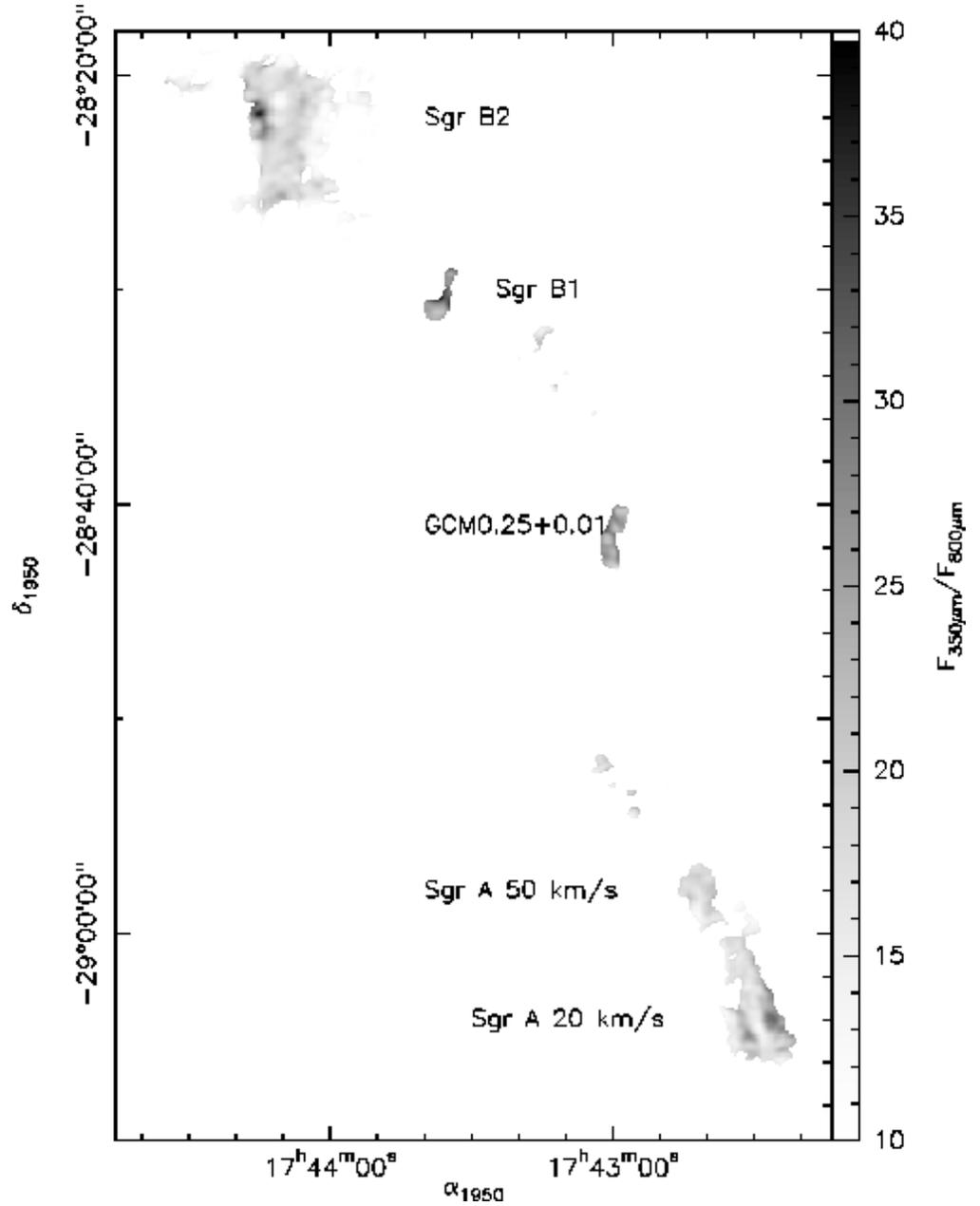}
\caption{Flux ratio F$_{350\micron}$/F$_{800\micron}$.  The 800 \micron\ map
is by Lis \& Carlstrom (1994).  The 350 \micron\ map has been smoothed to
30\arcsec\ resolution to match the 800 \micron\ map.  The flux ratio is
approximately 17 over most of the map.  A low spectral index is seen at the
cold, optically thick source Sgr B2 (N) at $17^h44^m10.3^s$,
$-28\deg 21\arcmin 17\arcsec$, and a high spectral index is observed at
M0.25+0.01 and Sgr B1 in the Dust Ridge.  The high spectral index to the
east of Sgr B2 and to the west of Sgr A is likely an artifact.}
\label{fig-h}
\end{figure}

The large coverage of the 350 \micron\ SHARC map and the 800 \micron\ map of
Lis \& Carlstrom (1994) allows an investigation of the spectral index over
the Galactic center region (Figure~\ref{fig-h}).  The mean value of
F$_{350 \micron}$/F$_{800 \micron}$ is approximately 17, which corresponds to
a spectral index of $\beta\ = 2.0$ for 30 K dust.  The flux ratio is low (12)
at Sgr B2 (N) due to its high optical depth.  The flux
ratio is high (27) at GCM0.25+0.01 and Sgr B1 in the Dust Ridge; the large
implied
$\beta\ = 2.5$ may be a signature of the lack of high-mass star formation
(Lis \& Menten 1998; Lis et al. 1998).

\acknowledgments

We thank T. Hunter for assistance in acquiring the data and F. Yusef-Zadeh and
D. Mehringer for providing the radio continuum maps.
The Caltech Submillimeter Observatory is supported by grant AST 96-15025 from
the NSF.

\end{document}